\documentclass{aa}
\usepackage{txfonts}

\begin{document}

\title{GRB 110709B in the Induced Gravitational Collapse paradigm}
\titlerunning{GRB 110709B}
\authorrunning{A.V. Penacchioni et al.}
\author{A.V. Penacchioni\inst{1,3}, R. Ruffini\inst{1,2}, C.L. Bianco\inst{1,2}, L. Izzo\inst{1}, M. Muccino\inst{1}, G.B. Pisani\inst{1,3}, J.A. Rueda\inst{1,2}}
\institute{Dip. di Fisica, Sapienza Universit\`a di Roma and ICRA, Piazzale
Aldo Moro 5, I-00185 Roma, Italy. E-mail: [ruffini;luca.izzo;marco.muccino;bianco;jorge.rueda]@icra.it
\and
ICRANet, Piazzale della Repubblica 10, I-65122 Pescara, Italy. 
\and
Universit\'e de Nice Sophia Antipolis, Nice, CEDEX 2, Grand Chateau Parc Valrose, E-mail: ana.penacchioni@icra.it }

\abstract{GRB 110709B is the first source for which \textit{Swift} BAT triggered twice, with a time separation of  $\sim$ 10 minutes. The first emission (Episode 1) goes from 40 s before the first trigger up to 60 s after it. The second emission (Episode 2) goes from 35 s before the second trigger to 100 s after it. [...] Within the Induced Gravitational Collapse (IGC) model, we assume the progenitor to be a close binary system composed of a core of an evolved star and a Neutron Star (NS). The evolved star explodes as a Supernova (SN) and ejects material that is partially accreted by the NS. We identify this process with Episode 1. The accretion process brings the NS over its critical mass, thus gravitationally collapsing to a BH. This process leads to the GRB emission, Episode 2. [...]
}
{We analyze the spectra and time variability of Episode 1 and 2 and compute the relevant parameters of the binary progenitor and the astrophysical parameters both in the SN and the GRB phase in the IGC paradigm. 
}
{We perform a time-resolved spectral analysis of Episode 1 by fitting the spectrum with a blackbody (BB) plus a power-law (PL) spectral model. We analyze Episode 2 within the Fireshell model, identifying the Proper-GRB (P-GRB) and simulating the light curve and spectrum. We establish the redshift to be $z=0.75$, following the phenomenological methods by Amati, by Yonetoku and by Grupe, and our analysis of the late X-ray afterglow[...].
}
{We find for Episode 1 a temperature of the BB component that evolves with time following a broken PL, with the slope of the PL at early and late times times $\alpha= 0$ and $\beta= -4 \pm 2$, respectively. The break occurs at $t= 41.21$ s. The total energy of Episode 1 is $E_{iso}^{(1)}= 1.42 \times 10^{53}$ erg. The total energy of Episode 2 is $E_{iso}^{(2)}=2.43 \times 10^{52}$ erg. We find at transparency a Lorentz factor $\Gamma \sim 1.73 \times 10^2$, laboratory radius of $6.04 \times 10^{13}$ cm, P-GRB observed temperature $kT_{P-GRB}= 12.36$ keV, baryon load $B=5.7 \times 10^{-3}$ and P-GRB energy of $E_{P-GRB}=3.44 \times 10^{50}$ erg. [...]
}
{We interpret GRB 110709B as a member of the IGC sources, together with GRB 970828, GRB 090618 and GRB 101023. The existence of the XRT data during the prompt phase of the emission of GRB 110709B (Episode 2) offers an unprecedented tool for improving the diagnostic of GRBs emission. }


\maketitle

\section{Introduction}\label{introduction and observations}

Of all the astrophysical processes currently being analyzed, few are more fundamental than the one presenting the coincidence of some Gamma-Ray Bursts (GRBs) with the explosion of a Supernova (SN). For this, the Induced Gravitational Collapse (IGC) paradigm was first introduced by \citet{Ruf2001} and further analyzed in \citet{Ruffini2007}, \citet{MG2008}, \citet{Jorge} and \citet{LucayJorge}. Recently, it has been evidenced that indeed this process can explain the coincidence between the SN and GRB emission, both from an observational and a theoretical point of view \citep{Izzo,Penacchioni2012}. 

In the IGC paradigm \citep{Ruf2001,Ruffini2007}, a binary system formed by an evolved star and a Neutron Star (NS) companion is considered as the progenitor. 

The IGC paradigm implies a well determined time sequence.
In a close binary system of a massive star in the latest phases of its thermonuclear evolution and a NS companion, the massive star undergoes a SN explosion.
The accretion of the early-SN material onto the NS companion leads the NS to its critical mass and consequently to its gravitational collapse to form a black hole (BH). 
The emission of a canonical GRB in the collapse to the BH takes place. A young NS  is born out of the SN explosion.
Finally, a SN emission is either observed or expected in association with the GRB, $\sim$ 10 days after the burst in the rest frame.  
We aim to find sources in which the data are of such a good quality to allow to see this complete sequence. 

The prototype for the IGC paradigm has recently been given in the analysis of GRB 090618 \citep{Izzo}, following the works of \citet{Jorge} and \citet{LucayJorge}. In this work we follow the same line and identify four different episodes in GRB 110709B. Episode 1 starts 40 s before the first trigger and lasts up to 60 s after it and is well fit by a blackbody (BB) plus a power-law (PL) spectral model . It corresponds to the trigger of the SN explosion of the compact core and its accretion onto the NS companion. We notice that the BB temperature decays with time following a broken power-law \citep{Ryde}. Episode 2 starts 35 s before the second trigger and lasts up to 100 s after it. It corresponds to the emission of the canonical GRB emitted in the formation of a BH. Episode 3 starts at 800 s all the way to $10^6$ s. It consists in a standard X-ray emission identified in all systems following the IGC paradigm (Pisani et al., in preparation). Episode 4 corresponds to the observation of the optical SN emission, observable after $T_{obs}=(1+z)T_{SN}$. In the present case, there is no evidence of an associated SN in the optical band. An explanation for this is given by \citet{Zaudereretal2012}, who classified GRB 110709B as dark and stated that its optical emission may have been absorbed by the host galaxy and/or the interstellar medium. 
The ensemble of these four episodes characterize the IGC scenario.

As an outcome, at the endpoint of the IGC scenario, a binary system represented by a NS (formed by the SN explosion) and a BH (formed after the GRB explosion) should be expected.

As in the case of GRB 101023, we do not know the cosmological redshift of GRB 110709B due to the lack of optical data.
Therefore, we infer it from phenomenological methods: 1) The Amati relation \citep{Amati(2006)},  2) the Yonetoku relation \citep{Yonetoku2004, Yonetoku2004b}, 3) the work of \citet{Grupe} and 4) the work by \citet{Penacchioni2012}, \citet{Ruffini2012} and by Pisani et al. (in preparation), which describe a scaling of the late X-ray emission of GRB 090618. In the case of GRB 111228, which we are currently analyzing, we find a striking coincidence between the values of the cosmological redshift determined by these methods for GRB 110709B.

In section \ref{observations} we report the observations of the two components of GRB 110709B by the different instruments, in space and on the ground.
In section \ref{data analysis} we reduce the \textit{Swift} data and perform a detailed spectral analysis of both Episode 1 and Episode 2. 
In section \ref{redshift estimation} we infer the redshift of the source using the four phenomenological methods mentioned above. 
In section \ref{radius} we determine the radius of the emitting region from the knowledge of the redshift and the BB flux of the first episode.
In section \ref{episode2} we give a brief description of the Fireshell model and we perform a deeper analysis of Episode 2 within this model, reproducing the light curve and the spectrum by a numerical simulation.
In section \ref{Nature of the Progenitor} we calculate the parameters of the binary progenitor leading to the IGC of the NS to a BH by the SN explosion. Details on the accretion rate onto the NS, total accreted mass, SN ejecta density, NS mass, and binary orbital period are obtained for selected values of the SN progenitor mass.
In section \ref{radio} we comment on the radio emission detected by EVLA \citep{GCN12190}.   
In section \ref{conclusions} we present the conclusions.

\section{Observations of GRB 110709B}\label{observations}

GRB 110709B has been detected by \textit{Suzaku} \citep{GCN12172} and \textit{Swift} \citep{GCN12122} satellites, and by ground-based telescopes like \textit{GROND} \citep{GCN12129} and \textit{Gemini} \citep{GCN12128}.

\begin{figure}
\centering
\includegraphics[height=0.8\hsize, angle=-90]{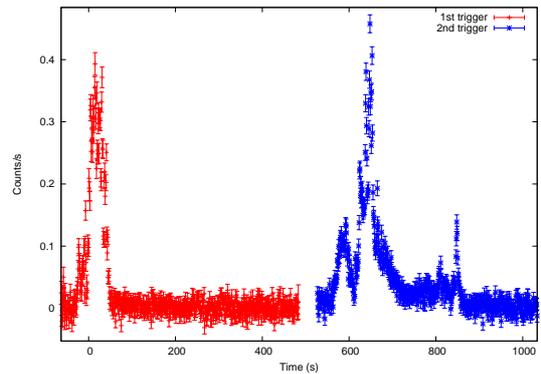}
\caption{BAT light curve of GRB 110709B, including both triggers. Here we can appreciate the time separation (about 10 minutes) between the first and the second trigger. The light curve is in the (15-150 keV) energy band. The time is relative to the first trigtime, of 331939966 s (in MET seconds). The second trigger was at 331940612 s in MET seconds.}
\label{LC_BAT1y2}
\end{figure}

\begin{figure}
\centering
\includegraphics[height=0.9\hsize, angle=-90]{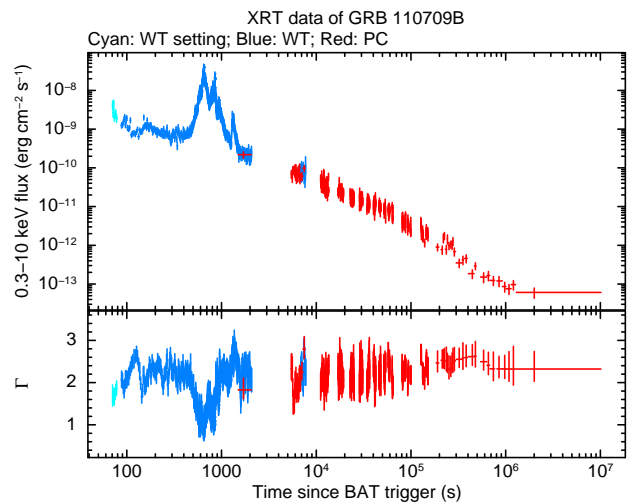}
\caption{Count light curve of GRB 110709B obtained from the \textit{Swift}-XRT detector, in the (0.3-10 keV) energy band. The time is relative to the first trigtime, of 331939966 s  (in MET seconds).}
\label{lc_XRT}
\end{figure}

The Burst Alert Telescope (BAT) on board \textit{Swift} triggered a first time at 21:32:39 UT (trigger N$^\circ$= 456967). The location of this event is RA= 164.6552, DEC= -23.4550. The light curve is composed of multiple peaks, with the whole emission extending up to 60 s after the trigger (see Fig. \ref{LC_BAT1y2}). What is most interesting is that there was another trigger at 21:43:25 UT (trigger N$^\circ$=456969), $\sim$ 11 minutes after the first trigger. The on-board calculated location is RA=164.647, DEC=-23.464. This time \textit{Swift} did not need to slew, because it was already pointing to that position. 
This second emission shows a bump that begins 100 s before the second trigger and lasts around 50 s, followed by several overlapping peaks with a total duration of about 40 s, and another isolated peak of 10 s of duration, 200 s after the second trigger. 
Fig. \ref{LC_BAT1y2} shows the complete BAT light curve and Fig. \ref{lc_XRT} shows the light curve taken by the X-Ray Telescope (XRT) in the 0.3-10 keV band. 

There have not been detections in the optical band by \textit{Swift} UVOT, which started to observe 70 s after the first BAT trigger \citep{GCN12157}. 
The observations with GROND at La Silla Observatory \citep{GCN12129} simultaneously in the g' r' i 'z' JHK, reveal two point sources within the 5'' .3 XRT error circle reported by \citet{GCN12122}. They suggest that one of them could be an afterglow candidate for GRB 110709B, although it is very faint. 

It has been suggested by \citet{Zaudereretal2012} that this source is an ``optically dark" GRB. The possible reasons for this are: 1) dust obscuration, 2) intrinsically dim event, and/or 3) high redshift (optical emission suppressed by Ly$\alpha$ absorption at $\lambda_{obs} \leq 1216 \AA \, (1+z))$. However, they rule out the possibility of a high redshift event due to the association with an optically detected host galaxy. Furthermore, they have inferred the optical brightness of the afterglow according to the standard afterglow synchrotron model \citep{GranotSari2002,Sarietal1999}, and from the non-detection in the optical-NIR wavelengths they find a very large rest frame extinction for GRB 110709B . This can explain the lack of detections in the optical band.

There have been detections in the radio band on several occasions by EVLA \citep{GCN12190}, revealing a single unresolved radio source within the XRT error circle, which rebrightened by a factor of 1.6 between 2.1 and 7 days after the burst. The location of the source is RA=10:58:37.114, DEC=-23:27:16.760.

\section{Data analysis}\label{data analysis}

\begin{figure}
\centering
\includegraphics[width=0.9\hsize]{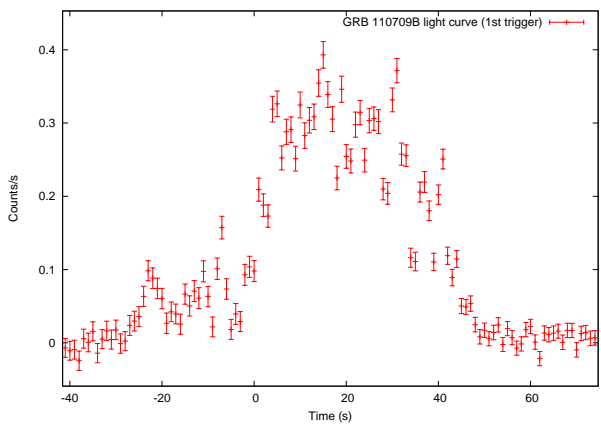}
\caption{Count light curve of Episode 1 of GRB 110709B obtained from the \textit{Swift}-BAT detector, in the (15-150 keV) energy band. The time is relative to the first trigtime, of 331939966 s (in MET seconds).}
\label{lc_1trigger}
\end{figure}

In the following we refer to the emission that goes from 40 s before the first BAT trigger to 60 s after it as Episode 1 (see Fig. \ref{lc_1trigger}). We call the emission going from 35 s before the second BAT trigger to 100 s after it as Episode 2. We make use of \textit{Swift} BAT data to perform the spectral analysis with XSPEC.

\subsection{Episode 1}\label{subsec:ep1}

\begin{table*}
\centering
\caption{Fit results of Episode 1 with five spectral models: BB, Band, BB+PL, PL and CutoffPL. The flux is in the energy band (15- 150) keV, in units of erg/cm$^2$/s.} 
\label{TABLATODO} 
\begin{tabular}{cccccc}
\hline
BB&Band&BB+PL&PL&CutoffPL\\
\hline
$kT=18.9 \pm 0.2$&$\alpha = -1.2 \pm 0.1$ & $kT =22 \pm 5$&$\gamma =1.37 \pm 0.02$&$\gamma = 1.1 \pm 0.1$\\
$K_{BB}= 0.95 \pm 0.01$&$\beta=$unconstr.&$K_{BB}=0.2 \pm 0.1$&$K_{PO}=2.0 \pm 0.2$&$E_0=196 \pm 68$\\
&$E_0=296 \pm 255$&$\gamma = 1.4 \pm 0.1$&&$K =0.8\pm 0.2$\\
&&$K_{PO}=2.2 \pm 0.8$&&&\\
\hline
Red$\chi^2 = 7.3$&Red$\chi^2 = 1.031$&Red$\chi^2 = 1.049$&Red$\chi^2 = 1.14$&Red$\chi^2 = 0.99$\\
56 DOF&54 DOF&54 DOF&56 DOF&55 DOF\\
Flux$= 7.52 \times 10^{-8}$&Flux$=8.99 \times 10^{-8}$&Flux$=8.96 \times 10^{-8}$&Flux$=9.08 \times 10^{-8}$&Flux$=8.93 \times 10^{-8}$\\
\hline
\end{tabular}
\end{table*}

\begin{figure}
\centering
\includegraphics[angle=-90,width=9.5cm]{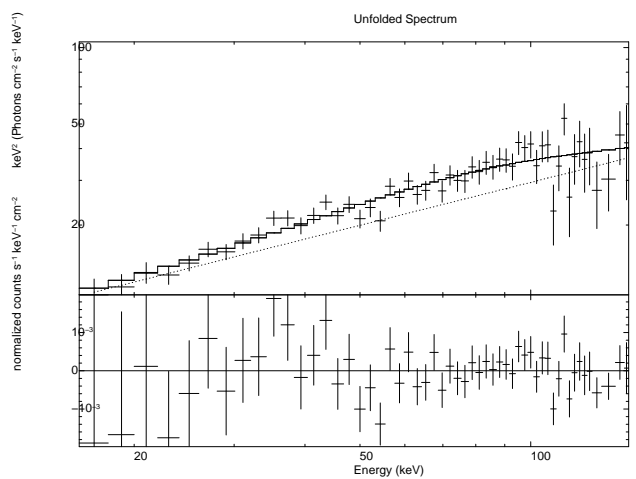}
\caption{Fit of Episode 1 with a BB+PL model. The parameters of the fit are: $kT=(22 \pm 5)$ keV, BB Amp$=0.2 \pm 0.1$, $\gamma= 1.4 \pm 0.1$, PL Amp$= 2.2 \pm 0.8$, Red-$\chi ^2= 1.049$ (54 DOF).}
\label{BB+PLEp1}
\end{figure}

\begin{figure}
\centering
\includegraphics[width=0.9\hsize]{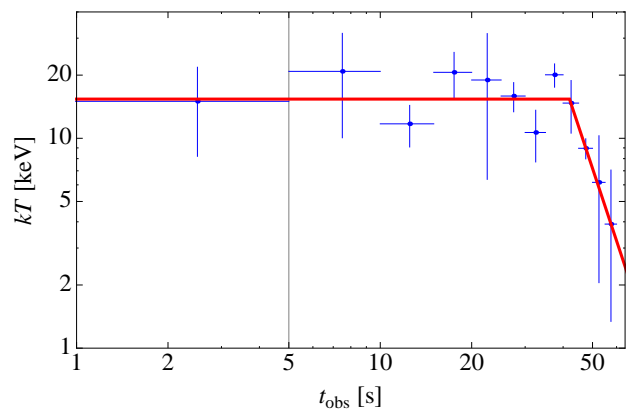}
\caption{Evolution of the kT component of the BB+PL model during Episode 1. The first data point corresponds to 5 s before the first BAT trigger. The vertical line corresponds to the trigger time. The time is in the observer frame. The temperature evolves in time following a broken power-law fit. There is a break at $t=41.21$ s. The indices of the PL are $\alpha= 0$ (consistent with a constant function) and $\beta= -4 \pm 2$, respectively. 
The presence of the BB, although smaller than in previous cases, is essential to have an acceptable fit.}
\label{broken powerlaw}
\end{figure}

We perform a time-integrated analysis to the whole Episode 1, using five different spectral models, namely BB, Band \citep{Band1993}, BB+PL, PL and CutoffPL. The results of the fits are shown in Table \ref{TABLATODO} and in Fig. \ref{BB+PLEp1}. The Band function is not well constrained, so we have excluded it in the following analysis. A statistical test shows that the best models are BB+PL ($\chi ^2=56.65$) and CutoffPL ($\chi ^2= 54.45$). Since the difference in the $\chi ^2$ between these two models is 2.2, the two models are statistically equivalent. So we discriminate between these two models based on the physical grounds expected from the IGC scenario. In this scenario, we expect a thermal emission from the expansion of the outer layers of the compact core SN progenitor. Thus, we have chosen the BB+PL model. We obtain a BB temperature $kT=(22 \pm 5)$ keV, a PL index $\gamma= 1.4 \pm 0.1$ and a $\chi ^2= 56.65$ (54 DOF). The flux of the BB component is $\sim 12 \%$ of the total flux. The total energy of Episode 1 is $E_{iso}^{Ep_1}=1.42 \times 10^{53}$ erg. The results of the fit are shown in Table \ref{TABLATODO}. Then we perform a time-resolved spectral analysis with a binning of 5 s fitting the same model and find that the temperature of the BB component follows a broken power-law, as mentioned in \citet{Ryde}, from 5 s before the trigger to 55 s after it (see Fig. \ref{broken powerlaw}). The broken power-law is indeed a constant function plus a simple power-law function. This is the same behavior as for the previously analyzed GRB 090618 \citep{Izzo} and GRB 101023 \citep{Penacchioni2012}. However, we notice that the temperatures for this GRB are lower. 
Nevertheless, the simultaneous presence of a BB and PL component is necessary in order to obtain an acceptable fit of the data (see Fig. \ref{BB+PLEp1}).

\subsection{Episode 2}\label{subsec:ep2}

\begin{figure}
\centering
\includegraphics[width=0.9\hsize]{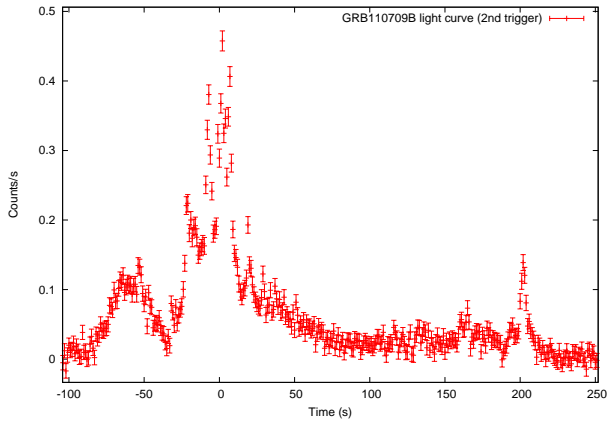}
\caption{Count light curve of Episode 2 of GRB 110709B obtained from the \textit{Swift}-BAT detector, in the (15-150 keV) energy band. The time is relative to the second trigtime, of 331940612 s (in MET seconds).}
\label{lc_2trigger}
\end{figure}

We also performed a time-integrated spectral analysis of Episode 2, whose light curve is shown in Fig. \ref{lc_2trigger}. This episode starts 35 s before the second trigger and last 135 s, until 100 s after the second trigger. We tried to fit the spectrum with the following spectral models: BB, PL, BB+PL, cutoffPL and Band (see Table \ref{fitep2}). We can easily discard the BB and Band models because in one case the Red $\chi^2$ is too high and in the other case there is an unconstrained parameter. As the PL and the BB+PL are nested models, we performed a statistical test to see which one is the best. We obtained a probability Prob=0.001 that the simpler model is better, so the BB+PL dominates over the PL. Then we have to compare between the BB+PL and the CutoffPL models. As they are not nested, we cannot apply the same test. So we chose the model that gives the lowest $\chi^2$. We concluded that the model that best fits Episode 2 is the cutoffPL model.

It is clear from the analogies with GRB 090618 and GRB 101023 that Episode 2 has all the characteristics of a canonical GRB. A difference between GRB 110790B and the already analyzed ones is that the separation between Episode 1 and Episode 2, $\sim 10$ min, is much bigger than previously, $\sim 50$ s. This remarkable time separation between the two episodes is an additional new fact to propose a different astrophysical origin of these two components.

We turn now to the crucial analysis of the determination of the cosmological redshift of Episode 2. 

\begin{figure}
\centering
\includegraphics[width=0.7\hsize, angle=-90]{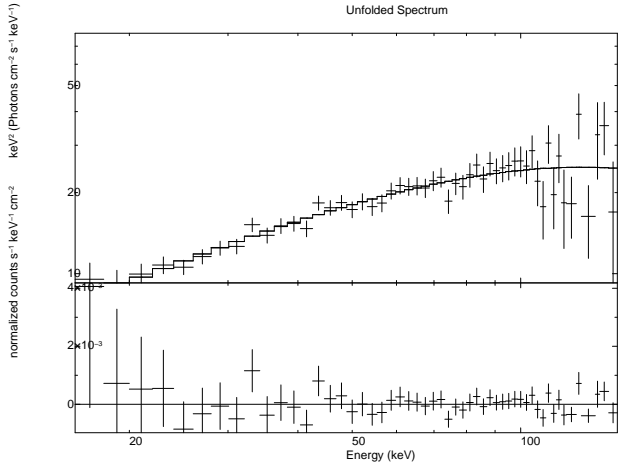}
\caption{Fit of Episode 2 with a Cutoffpl model. The photon index is $\gamma = 1.0 \pm 0.1$, the normalization constant is $0.5 \pm 0.1$, the cutoff energy is $E_0 = 132 \pm 31$ and the reduced chi squared of the fit is Red-$\chi ^2= 0.77$ (55 DOF).}
\label{PL}
\end{figure}  

\begin{table*}
\centering
\caption{Spectral fit of the whole Episode 2 with different models: BB, PL, BB+PL, cutoffPL and Band. The flux corresponds to the (15 -150) keV energy range. The model that best fits the data is the cutoffPL. We can easily see that the BB and Band models are not good. As the PL and BB+PL are nested models, we performed a statistical test to see which was the best one. We obtained a probability  Prob=0.001, indicating that the BB+PL dominates over the PL model. Then, between BB+PL and CutoffPL (that are not nested, then we cannot apply the same test), we took the CutoffPL as the best model because it gives the lowest $\chi^2$. Note that the models with which we fit the data are those defined in the XSPEC's Manual: http://heasarc.nasa.gov/xanadu/xspec/xspec11/manual/manual.html } 
\label{fitep2} 
\begin{tabular}{ccccc}
\hline
 BB& PL &BB+PL & CUTOFFPL & BAND  \\
 \hline
 $kT$ [keV]$=17.5 \pm 0.2$ & $\gamma=1.46 \pm 0.02$ &$kT$ [keV]$=20 \pm 3$ & $\gamma=1.0 \pm 0.1$ & $\alpha=-1.0 \pm 0.1$\\
 $K_{BB}= 0.661 \pm 0.009$  & $K_{PO}=2.1 \pm 0.2$ & $K_{BB}=0.16 \pm 0.06$  & $E_0= 132 \pm 31$ & $\beta=$ unc\\
   &  & $\gamma=1.5 \pm 0.1$& $K=0.5 \pm 0.1$ & $E_0= 142 \pm 42$ \\ 
   &  & $K_{PO}=2.3 \pm 0.8$ & & $K=0.0048 \pm 0.0008$ \\
   \hline
 Red $\chi ^2=7.16$ & Red $\chi ^2=1.109$ & Red $\chi ^2=0.78$ & Red $\chi ^2=0.77$& Red $\chi ^2=0.79$\\
 DOF= 56 &  DOF= 56 &  DOF=54 &  DOF= 55 &  DOF= 54\\
 Flux= $5.2 \times 10^{-8}$ & Flux= $6.52 \times 10^{-8}$& Flux =$6.35 \times 10^{-8}$& Flux= $2.43 \times 10^{-8}$& Flux= $6.36 \times 10^{-8}$\\
\hline
\end{tabular}
\end{table*}

\section{Cosmological redshift determination}\label{redshift estimation}

We used four phenomenological methods to constrain the redshift of the source, based on different relations, detailed below.

\subsection{$N_H$ column density}

We first tried to get an upper limit for $z$ following the work of \citet{Grupe}. They consider a relation between the absorption column density in excess of the galactic column density, given by $\Delta N_H=N_{H,fit}-N_{H,gal}$ and the redshift $z$, through the equation

\begin{equation}
\label{formulaGrupe}
log(1+z) < 1.3-0.5[log(1+\Delta N_H)].
\end{equation}

We calculated $N_{H,gal}$ from the radio map of the galaxy in the Lab Survey website\footnote{http://www.astro.uni-bonn.de/$\sim$webaiub/english/tools$\_$labsurvey.php} by entering the coordinates of the GRB (RA=164.64, DEC=-23.46). We obtained $N_{H,gal}=10.5 \times 10^{20}$ cm$^{-2}$. 

To obtain the value of $N_{H,fit}$ we took the XRT data from 2000 s to $10^6$ s after the first BAT trigger and fitted the model \textit{phabs*po} using the program XSPEC. The XRT data were reduced by the xrtpipeline software, version 0.10.4, which is part of the HEASOFT package, version 6.12. We use the standard response matrix swxpc0to12s6$\_$20010101v013.rmf for the PC mode data. The model \textit{phabs} represents the photoelectric absorption
\begin{equation}
M(E)=e^{-n_H \sigma(E)},
\end{equation}
where $n_H$ is the equivalent hydrogen column density (in units of $10^{22}$ cm$^{-2}$) and $\sigma(E)$ is the photoelectric cross section, not including Thompson scattering. 
We obtained a value of $N_{H,fit}=71.76 \times 10^{20}$ cm$^{-2}$. Using these values in (\ref{formulaGrupe}) we obtained an upper limit for the redshift of $z<1.35$. 

\subsection{Amati Relation}

We also tried to determine the redshift of Episode 2 through the Amati relation \citep{Amati(2006)}, that relates the isotropic energy $E_{iso}$ of the GRB to the peak Energy in the rest frame $E_{p,i}$ of the $\nu F_{\nu}$ spectrum \citep{Amati2009}. 
The analytical expression of $E_{iso}$ is

\begin{equation}
\label{Eiso}
E_{iso}=\frac{4 \pi d_L^2}{(1+z)} S_{bol},
\end{equation}
where $d_L^2$ is the luminosity distance, $z$ is the redshift and $S_{bol}$ is the bolometric fluence, related to the observed fluence in a given detection band ($E_{min}$, $E_{max}$) by

\begin{equation}
S_{bol}=S_{obs}\frac{\int^{10^4 / (1+z)}_{1/ (1+z)} E \phi(E) dE}{\int ^{E^{max}}_{E^{min}} E \phi(E) dE}.
\end{equation}
Here, $\phi$ is the spectral model considered for the spectral data fit; in this case a Band model \citep{Band1993}, composed of two smoothly connected power-laws. $E_{p,i}$ is related to the peak energy $E_p$ in the observer frame by
\begin{equation}
E_{p,i}=E_p (1+z).
\end{equation}
The peak energy is the energy at the peak of the $\nu F_{\nu}$ spectrum. It can be written as $$E_p=E_0(2+\alpha),$$ where $E_0$ is the energy at which the two power-laws intersect and $\alpha$ is the slope of the low-energy power-law, according to the Band model. 

We calculate the luminosity distance $d_{L}$, as given by the standard cosmological model

\begin{equation}
d_L= \frac{c}{H_0}(1+z) \int^z_0{\frac{dx}{\sqrt{\Omega_m(1+x)^3+\Omega_{\Lambda}}}},
\end{equation}
 where the Hubble constant is $H_0 = 70$ km s$^{-1}$ Mpc$^{-1}$, $\Omega_m=0.27$, $\Omega_{\Lambda}=0.73$ and $c$ is the speed of light. 

\begin{figure}
\centering
\includegraphics[width=0.9\hsize, angle=0]{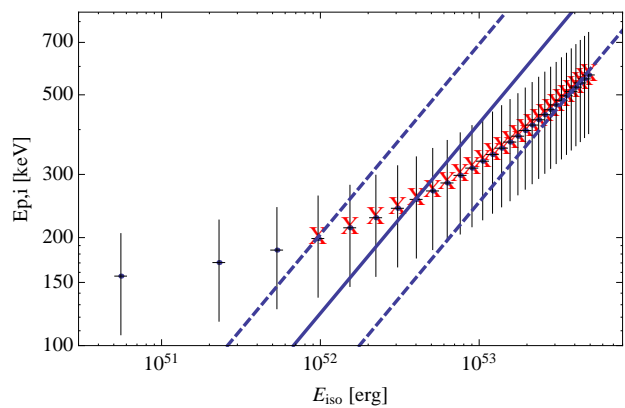}
\caption{Amati relation (solid line) with its $1-\sigma$ uncertainties (doted lines) and peak energy $E_{p.i}$ vs. $E_{iso}$ for GRB 110709B, for different values of the redshift, from 0.1 to 3, at steps of 0.1. We can see that the data matches the theoretical function within $1-\sigma$ for $z>0.4$.} 
\label{Amati}
\end{figure}

Following the same procedure as described in \citep{Penacchioni2012}, we calculated $E_{iso}$ and $E_{p,i}$ for different values of $z$, from 0.1 to 3, at steps of 0.1. Fig. \ref{Amati} shows that the relation is satisfied for values of $z>0.4$. This puts a lower limit to the estimation of the redshift. 

\subsection{Yonetoku Relation}

We finally obtained a range of possible redshifts by using the Yonetoku relation \citep{Yonetoku2004}. This relation, also known as the $E_p$ - Luminosity relation ($E_p$ - L) , connects the observed isotropic luminosity $L$ in units of $10^{52}$ erg s$^{-1}$ with the peak energy $E_p (1+z)$ in the rest frame of the GRB. It is valid for values of $E_p$ between 50 and 2000 keV, and a luminosity range of $10^{50} - 10^{54}$ erg s$^{-1}$.

The best fit function for the $E_p$ - L relation is

\begin{equation}
\frac{L}{10^{52}\rm{erg\, s}^{-1}}=(2.34^{+2.29}_{-1.76}) \times 10^{-5} \left[\frac{E_p(1+z)}{1~ \rm{keV}}\right]^{2.0 \pm 0.2}
\end{equation} 

The peak luminosity and the peak energy are calculated by integrating within a 1 s interval around the most intense peak of the light curve, because this is a better distance indicator than the burst average luminosity. However, we took a 10s interval around the most intense peak in order to better constrain the value of the parameters (i.e., to increase the number of photons in the spectrum and obtain an error which is smaller than the value of the parameters). The peak luminosity in the rest frame (with the proper K-correction) can be calculated as

\begin{equation}
L=4 \pi d_L^2 F_{bol},
\end{equation}
where 
\begin{equation}
F_{bol}=P_{obs} \frac{\int^{10000/(1+z)}_{1/(1+z)} E N(E) dE}{\int^{E_{max}}_{E_{min}} N(E) dE} 
\end{equation}
is the energy flux and $P_{obs}$ is the photon flux.
 
\begin{figure}
\centering
\includegraphics[width=0.8\hsize, angle=0]{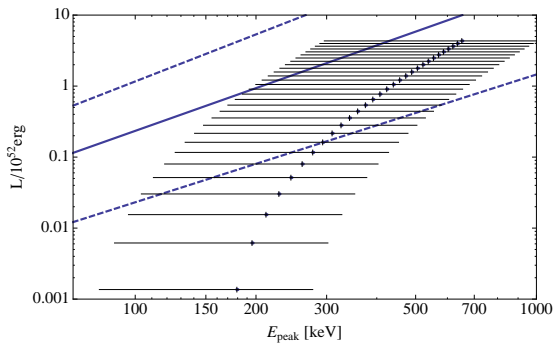}
\caption{Yonetoku relation (solid line) with its $1-\sigma$ uncertainties (doted lines) and peak Luminosity vs. $E_{p.i}$ for GRB 110709B, for different values of the redshift, from 0.1 to 3, at steps of 0.1. We can see that the data matches the theoretical function within $1-\sigma$ for $z>0.6$.}
\label{Yonetokurelation}
\end{figure}

 Fig. \ref{Yonetokurelation} shows the Yonetoku relation (solid line) with its uncertainties (dotted lines), and the values of $L$ and $E_{p,i}$ for each value of $z$, from 0.1 to 3, at steps of 0.1. We see that the Yonetoku relation is satisfied within $1 \sigma$ for values of the redshift $>0.7$, consistent with the results obtained with the Amati relation. 
 
 In conclusion, if we put together the three methods, we have a range of possible redshifts of $0.7 < z <1.35$. 
 
\subsection{Estimate of the redshift using the X-ray afterglow}\label{sec:afterglow}

We already presented in \citep{Penacchioni2012} a method to estimate the redshift of GRB 101023 by comparing its X-ray light curve to the one of GRB 090618, of known redshift ($z=0.54$). Here we rescale the X-ray light curve of GRB 090618 as if it was seen at different redshifts and plot it together with GRB 110709B light curve, looking for the values of $z$ for which these light curves overlap at late times. We find a remarkable consistency between this method and the phenomenological methods already mentioned. 

In oder to compare in a common rest frame the two emissions from the GRBs, we apply the following operations only to GRB 090618:  

1) determination of the starting time $T_{start}$ of the late decay emission,

2) spectral analysis of this emission with an absorbed power-law model,

3) extrapolation of this spectral model in a common cosmological rest-frame energy range and, consequently, rescaling of GRB 090618 light curve for the different energy ranges,

4) cosmological correction for the arrival time by taking into due account the different scaling due to cosmological redshift, and

5) correction of the observed flux by changing the redshift of GRB 090618.

A detailed description of the method will be given in a forthcoming publication (Pisani et al., in preparation).

\begin{figure}
\centering
\includegraphics[width=0.9\hsize]{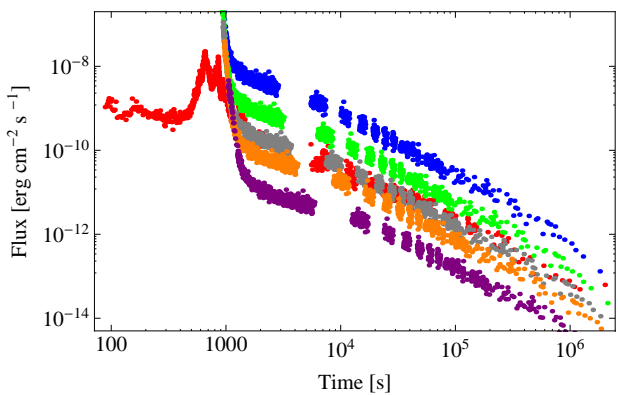}
\caption{Plot of GRB 090618 seen as if it was at different redshifts: 0.2 (blue), 0.4 (green), 0.7 (grey), 1.0 (orange) and 2.0 (purple). The red light curve corresponds to GRB 110709B. We can see that it lies between the green and the orange light curves, indicating that the redshift must be between 0.4 and 1.0.}
\label{afterglow2}
\end{figure}

In this way we compare directly both light curves for different redshifts of GRB 090618. Fig. \ref{afterglow2} shows GRB 090618 light curve seen as if the source was located at different redshifts: 0.2 (blue), 0.4 (green), 0.7 (grey), 1.0 (orange) and 2.0 (purple). The red light curve corresponds to GRB 110709B. We can see that it lies between the green and the orange ones. A more accurate scaling of the late X-ray afterglow suggests a redshift of $z=0.75$ for this source.

\begin{figure}
\centering
\includegraphics[width=0.9\hsize]{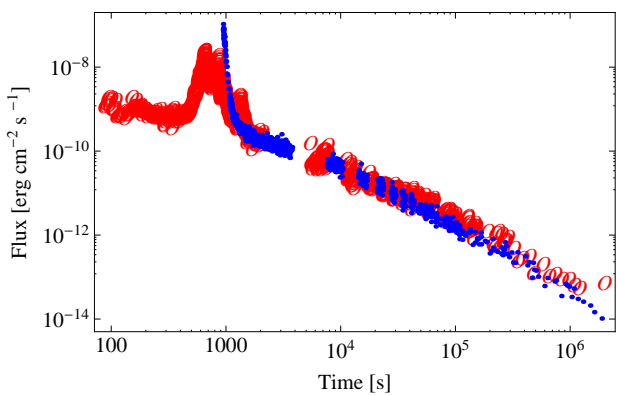}
\caption{GRB 090618 light curve (blue) as if it was seen at $z=0.75$ together with GRB 110709B light curve (red). There is an excellent superposition in the late decay assuming a temporal shift of GRB 090618 of $T_*=800$ s to the right to have the coincidence of the steep decays of the light curves.}
\label{afterglow1}
\end{figure}

Fig. \ref{afterglow1} shows the superposition of GRB 110709B and GRB 090618 light curves in the observer frame, as if they were located at a redshift $z=0.75$. 

There is however a second aspect which is due to the peculiarity of the turn-on $T_0$ of the XRT detector. At the time BAT triggered for the second time, XRT was already pointing at the source and was able to detect the emission at very early times, making this GRB probably the first for which XRT has the earliest detection up to date. We need to shift GRB 110709B light curve in order to make the early steep decays (originating in the prompt in our interpretation) coincide. This is done by adding a time $T_*=+800$ s to GRB 090618 light curve. The superposition is very good. In this way we also make the early decays coincide. This factor is arbitrary, but we need to include it because GRB 110709B XRT light curve presents many spikes at the beginning, which according to our interpretation correspond not to the steep decay of the X-ray light curve but to the prompt emission.

In the case of GRB 110709B, thanks to the fact that XRT was already active and collecting data at the time of the second BAT trigger, we were able to follow the behavior of the whole GRB emission of Episode 2. This is a key point to our understanding of GRB 110709B, since only in very few cases XRT had a response during the early emission. 

In the big flare at $\sim 1000$ s after the first BAT trigger we notice a strong correlation between the emission in X-rays and in $\gamma$-rays. We identify this emission as the prompt emission of Episode 2. After this prompt phase the traditional plateau phase is observed. After the plateau phase, there is the late decay phase in the X-ray light curve following a power-law behavior which has already been observed in other sources (i.e., GRB 101023, GRB 090618, GRB 111228). We study this decay in the IGC paradigm, and consider the possibility that it is produced by the early emission of the newly-born NS. It is interesting to notice that in GRB 110709B the typical flare in X-rays just preceding the plateau phase and following the prompt emission is not observed. This X-ray emission usually occurs without any associated $\gamma$-ray emission, since the data is usually below the BAT threshold. In the present case, it is conceivable that the flaring indeed occurred during some of the gaps of $\sim 4000$ s in which there is no data due to Earth occultation. 

We can then distinguish two types of flares in the X-ray light curve. The first type occurs at early times, previous to the steep decay, and belongs to the prompt emission. This flares can be seen in X-rays only when XRT starts its detection at early enough times, e.g., when the satellite was already pointing at a region near the burst position and did not need much time to slew. The light curve in X-rays generally follows the trend of the light curve in $\gamma$-rays. The second type of flares occurs at later times, just preceding the plateau phase. This flares are seen only in X-rays since their photon flux is much lower than the BAT threshold. In the ICG paradigm, we interpret this flares as possible indicators of the breakout of the SN. 

We are currently analyzing more sources in the catalogue by \citet{Grazia} to look for these three very distinct phases, i.e., the flares in the prompt emission, the flares in the afterglow and the late decay after the plateau, each of them having a different physical origin within the IGC paradigm.

\section{Episode 1: radius of the emitting region}\label{radius}

With the knowledge of the redshift and the parameters of the fit with a BB + PL model, we computed the isotropic energy of the whole Episode 1, $E_{iso}^{(1)}= 1.42 \times 10^{53}$ erg.

\begin{figure}
\centering
\includegraphics[width=0.9\hsize]{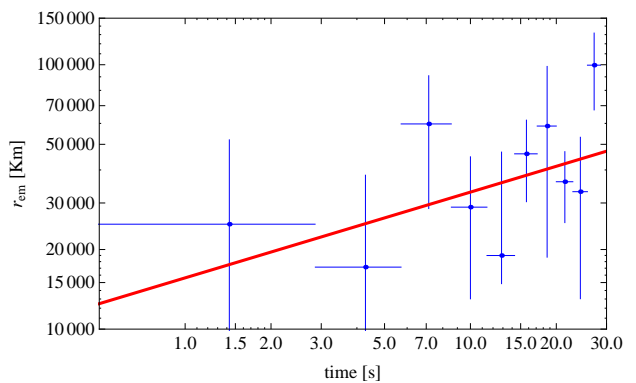}
\caption{Radius of the emitting region as a function of time (in the cosmological rest frame), corresponding to Episode 1. The radius increases with time following a power-law $a t^b$, with $a= (1.5 \pm 1.2)\times 10^4$ km/s and $b=0.32 \pm 0.27$.}
\label{Rem}
\end{figure}

With the energy flux of the BB component $\phi_{BB}$ as a function of time from the time-resolved spectral analysis and the luminosity distance $d_L$, we can compute the value of the radius of the emitter in cm (we then express it in km in Fig. \ref{Rem}) through

\begin{equation}
r_{em}=\sqrt{\frac{\phi_{BB}}{\sigma T^4}} \frac{d_L}{(1+z)^2}.
\end{equation}
Here $\phi_{BB}$ is the BB flux in units of $erg ~ cm^{-2} s^{-1}$, $\sigma=5.6704 ~ erg ~ cm^2 s^{-1} K^{-4}$ is the Stefan-Boltzmann constant and $d_L$ is the luminosity distance in cm.

The best fit of the expanding radius is
\begin{equation}
\label{r}
r(t)=a t^b,
\end{equation}
where $a= (1.5 \pm 1.2) \times 10^4$ km s$^{-b}$ and $b=0.32 \pm 0.27$ (see Fig \ref{Rem}). 

We associate the BB component to the expansion of the ejected material, while the power-law is associated (as we interpret from the IGC paradigm) with the accretion of part of this material onto the NS companion.

\section{Analysis of Episode 2 in the Fireshell model}\label{episode2}

We recall that the Fireshell model \citep{Damour75,RuffiniSalmonson2000,Ruffini2001,Ruffini2010b} is an alternative to the Fireball model, first proposed by \citet{CavalloRees}, \citet{Goodman} and \citet{Paczynsky}.
We assume, within the Fireshell model, that all GRBs originate from the gravitational collapse of a star approaching asymptotically the formation of a Kerr-Newmann BH \citep{The Kerr Spacetime: Rotating Black Holes in General Relativity}. An electric field $E$ is created just outside the collapsing core and in between the expanding outer shells that act as a capacitor \citep{Preparata}. This electrical field grows until it reaches a critical value, $E_c=m^2 c^3/\hbar e$. At this time, vacuum polarization occurs, leading to pair creation at the expenses of the gravitational energy. An optically thick $e^{\pm}$ plasma forms with total energy $E^{e^{\pm}}_{tot}$ in the range $10^{49}$ - $10^{54}$ erg. The $e^{\pm}$ plasma reaches thermal equilibrium on a timescale of $10^{-12}$ s \citep{Aksenov2007}. Being optically thick, the plasma self-accelerates due to its internal radiation pressure \citep{Ruffini1999,Ruffini1999b}. After an early expansion in vacuum, the $e^{\pm}$-photon plasma engulfs the baryonic matter $M_B$ of the outer shells and reaches thermal equilibrium with it. The baryonic matter is described by the dimensionless parameter $B = M_B c^2/E^{e^{\pm}}_{tot}$. $B$ must be less than $10^{-2}$, otherwise there will not be any relativistic expansion \citep{RuffiniSalmonson2000}. The optically thick fireshell composed by $e^{\pm}$-photon-baryon plasma self-accelerates to ultrarelativistic velocities, finally reaching the transparency condition. A flash of radiation is then emitted. This is the P-GRB \citep{Ruffini2001a}. The amount of energy radiated in the P-GRB is only a fraction of the initial energy $E^{e^{\pm}}_{tot}$. The remaining energy is stored in the kinetic energy of the optically thin baryonic and leptonic matter fireshell that expands ballistically and starts to slow down due to the inelastic collisions with the Circumburst Medium (CBM). This interaction gives rise to a multi-wavelength emission, the extended afterglow \citep{Ruffini2001a}. We can estimate the characteristic inhomogeneities of the CBM by fitting the luminosity of the X-ray source and imposing the fully radiative condition in the collision between the ultra relativistic baryonic shell and the clouds of the Interstellar Medium (ISM). The complete analytic solution has been developed in \citet{BiancoRuffini2004,BiancoRuffini2005a,BiancoRuffini2005b}, together with the analytic expression of the Surfaces of Equal arrival Time of the photons at the detector (EQTS). The afterglow presents three different regimes: a rising part, a peak and a decaying tail. We therefore define a ``canonical GRB'' light curve with two sharply different components: 1) the P-GRB and 2) the extended afterglow. What is usually called ``Prompt emission'' in the current GRB literature mixes the P-GRB with the raising part and the peak of the extended afterglow \citep{RuffiniBrazilianschool}. The spectrum of the extended afterglow is initially assumed to be thermal in the comoving frame of the expanding shell. Recently, after the analysis of some highly energetic sources observed by \textit{Swift} and \textit{Fermi} satellites, this assumption of a pure comoving thermal spectrum has been relaxed and a phenomenological modification by a power-law of the low energy spectral slope has been introduced \citep{Patricelli}.
The observed non thermal spectrum shape is due to a double convolution of thousands of instantaneous comoving spectra, with different temperatures and different Doppler factors, over both the EQTS and the observation time \citep{RuffiniIJMPD}. 

Having fixed the value of the redshift to $z=0.75$, we started the analysis of Episode 2 within the Fireshell model. We first looked for the P-GRB during the first bump of Episode 2 (from 100 to 40 s before the second trigger) by fitting the data with a BB + PL model. We selected several time intervals as the P-GRB during the first bump of Episode 2, but in some cases the fits were not good. In some other cases, to reproduce the ratio between the P-GRB energy and the total energy we needed to consider a baryon load $B>10^{-2}$ (which has no sense within the Fireshell model) and, in other cases, there was a discrepancy between the observed temperature and the one given by the simulation. Thus we concluded that this bump should belong to Episode 1. The reason why we do not find a strong thermal signature in this bump is that Episode 1 starts $\sim 10$ minutes before the beginning of the bump and the temperature of the BB component decreases very rapidly following a power-law in the first seconds of emission.Consequently, after such a long time we do not expect to find any signature of a BB from Episode 1. 

We finally selected the P-GRB as the 9 s from 35 to 26 s before the second trigger, and the following emission from -26 to 100 s as the afterglow. Table \ref{P-GRBEiso} shows the parameters of the fit. We calculated a P-GRB energy of $E_{P-GRB}=3.44 \times 10^{50}$ erg and an isotropic energy of $E_{iso}=2.43 \times 10^{52}$ erg.

\begin{table}
\centering
\caption{Fit of the P-GRB and the afterglow of GRB 110709B, Episode 2. The P-GRB is well fit with a BB model, while the whole Episode 2 is best fit by a CutoffPL model. From this fit and the value of the redshift we are able to calculate $E_{iso}$ and $E_{P-GRB}$. } 
\label{P-GRBEiso} 
\begin{tabular}{ccc}
\hline\hline
 Parameter & P-GRB& P-GRB+Afterglow  \\
 \hline
 $kT$ [keV] &$14 \pm 1$ &  \\
 BB Amp  & $0.30 \pm 0.02$ & \\
 $\gamma$ &  & $1.03 \pm 0.1$\\
 PL Amp &  & $0.5 \pm 0.1$\\
 Red $\chi ^2$ & $1.448$ (56 DOF) & $0.77$ (55 DOF) \\
Energy Flux & $2.413 \times 10^{-8}$ & $6.34 \times10^{-8}$\\
($15-150$ keV) & & \\ 
$[$erg cm$^{-2}$ s$^{-1}]$ & & \\
Energy [erg]& $3.44 \times 10^{50}$ & $2.43 \times 10^{52}$ \\
\hline
\end{tabular}
\end{table}

We inserted these values of the energies into our numerical code and calculated the value of the baryon load, $B = 5.7 \times 10^{-3}$. We simulated the light curve and the spectrum, obtaining, at the transparency point, a Laboratory Radius $r_{tr}=6.04 \times 10^{13}$ cm, a gamma Lorentz factor $\Gamma= 1.73 \times 10^2$ and a P-GRB observed temperature (after cosmological correction) $kT=12.36$ keV.

\begin{figure}
\centering
\includegraphics[width=0.5\hsize, angle=-90]{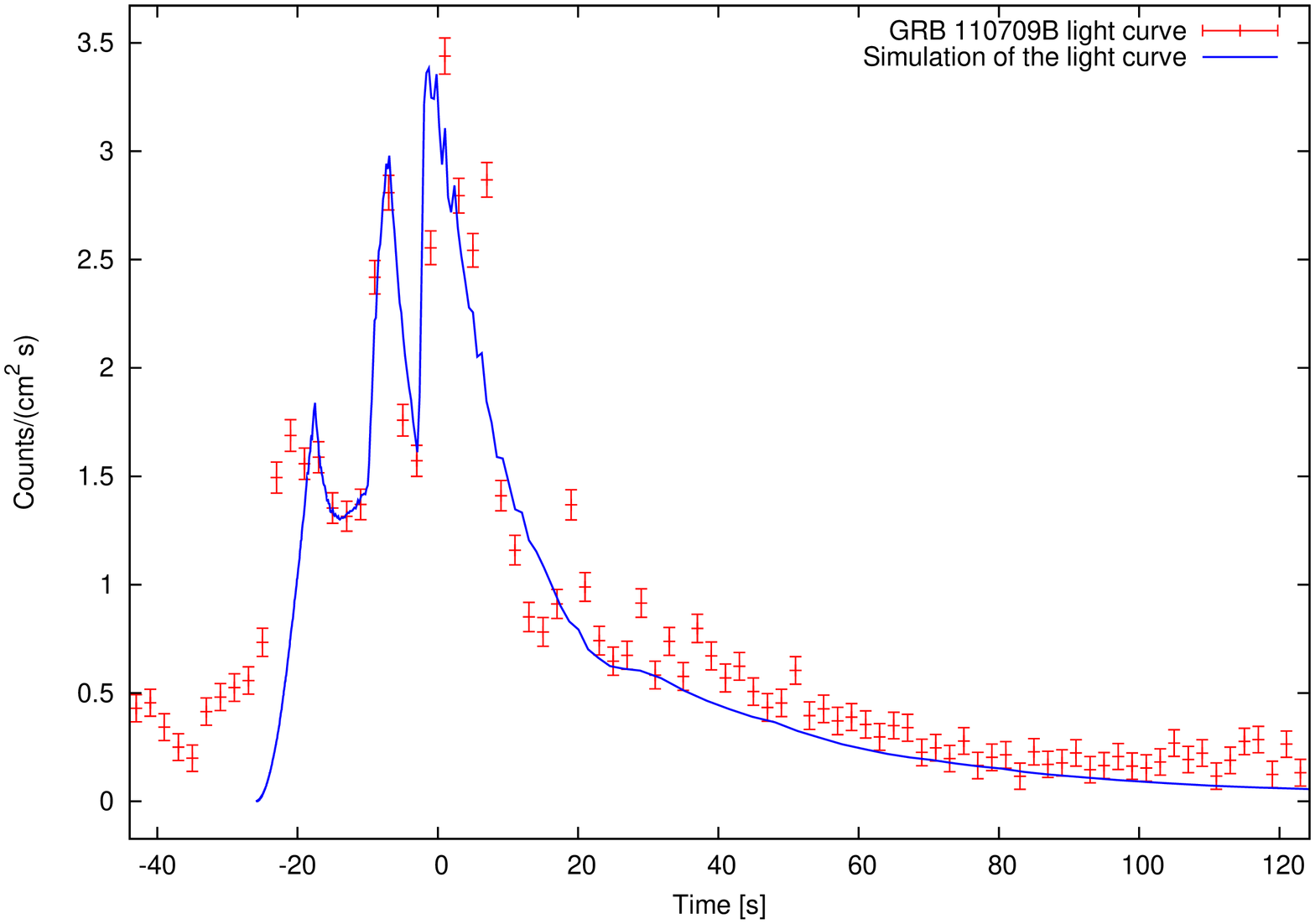}
\includegraphics[width=0.75\hsize, angle=0]{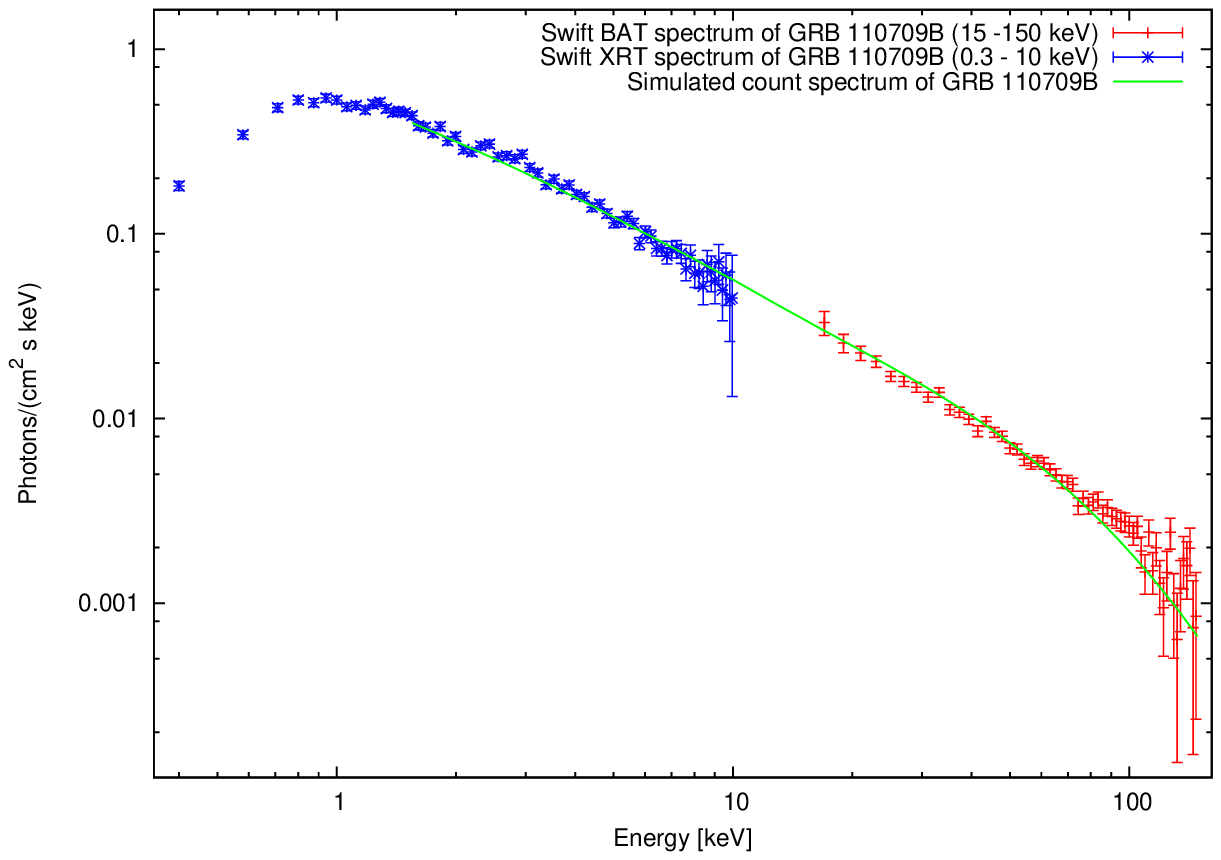}
\caption{Simulation of the BAT (a) light curve and (b) spectrum of Episode 2. We included XRT data in the fit of the spectrum to show that the slope predicted in the fireshell model is in agreement with the slope of the X-ray spectrum.}
\label{Sim}
\end{figure}

\begin{figure}
\centering
\includegraphics[width=0.5\hsize, angle=-90]{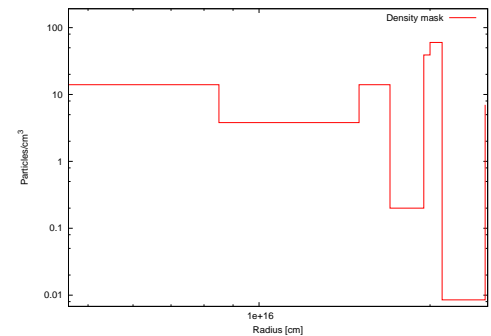}
\caption{Particle density of the ISM clouds as a function of the distance. The mean density is $76$ part/cm$^{3}$.}
\label{Densitymask}
\end{figure}

Figs. \ref{Sim}a and \ref{Sim}b show the simulation of the light curve and the spectrum of Episode 2, respectively. The photon index of the XRT and BAT spectra are in agreement with that predicted by the simulation. Details of this calculation will be given in a forthcoming letter (Penacchioni et al., in preparation). Fig. \ref{Densitymask} shows the density mask of the ISM, i.e., the density of particles of the interstellar clouds as a function of the distance to the center of the BH. This density has to be interpreted as an effective density because fragmentation may occur in the expanding shell \citep{Ruffini2007,Dainotti}. 

\section{Nature of the Progenitor}\label{Nature of the Progenitor}

Following the works of \citet{Jorge} and \citet{LucayJorge}, we suggest for the origin of GRB 110709B a binary system formed by a massive evolved star on the verge of a SN explosion and a NS. The early-SN material expanding at non-relativistic velocities is then accreted by the NS companion at times larger than $t_{0, \rm accr}$, when the material reaches the NS gravitational capture region. The emission observed in Episode 1 is associated to this early-SN evolution, identified with the thermal component, and accretion process onto the NS, possibly related to the non-thermal component. The NS reaches in a time $t_{0, \rm accr}+\Delta t_{\rm accr}$ the critical mass and gravitationally collapses to a black hole, emitting the GRB seen in Episode 2. We assume the critical mass of a non-rotating NS $M_{\rm crit}=2.67 M_{\odot}$ as given by \citet{Belvedere}. 

The amount of material that reaches the NS gravitational capture region
\begin{equation}
R_{\rm cap}(t)=\frac{2GM_{\rm NS}(t)}{v_{\rm ej,rel}^2(t)}
\end{equation}
per unit time is given by (see \citet{Jorge} and \citet{LucayJorge})

\begin{equation}\label{Mdot}
\dot{M}(t) = \pi \rho_{\rm ej}(t) v_{\rm ej,rel}(t) R_{\rm cap}^2(t),
\end{equation}
where $R_{\rm cap}$ is measured from the NS center.

In these expressions, $\rho_{\rm ej}(t)=3M_{\rm ej}(t)/(4 \pi r_{\rm ej}^3(t))$ is the density of the ejecta, 
\begin{equation}\label{Mej}
M_{\rm ej}(t)=M_{\rm ej}(0)-M(t)
\end{equation}
is the total available mass to be accreted by the NS, $M_{\rm NS}(t)$ is the NS mass, and $v_{\rm ej,rel}(t)$ is the velocity of the ejecta relative to the NS
\begin{equation}
v_{\rm ej,rel}(t)=\sqrt{v_{\rm orb,NS}^2(t)+v_{\rm ej}^2(t)}.
\label{v}
\end{equation} 
In Eq. (\ref{Mej}), $M_{\rm ej}(0)$ is the given initial mass of the ejecta (just at the beginning of the accretion process); we choose different values for it in Table \ref{Tablaprog}. $M(t)$ is the mass of the ejecta that is lost because it passes through the capture region of the NS.

The actual mass accretion rate onto the NS, $\dot{M}_{\rm accr}(t)$, is a fraction $\eta_{\rm accr} \leq 1$ of Eq. (\ref{Mdot}), i.e. 

\begin{equation}
\dot{M}_{\rm accr}(t)=\eta_{\rm accr} \dot{M}(t),
\end{equation}
where $\eta$ is the accretion efficiency onto the NS. So, there is an amount of material per unit time $\dot{M}_{out}(t)=(1- \eta_{\rm accr}) \dot{M}(t)$ not accreted by the NS.

In Eq. (\ref{v}), $v_{\rm orb,NS}(t)=\sqrt{G(M_{\rm prog}+M_{\rm NS}(t))/a}$ is the orbital velocity relative to the SN core progenitor, $a$ is the separation distance between the NS and the SN core progenitor, and
\begin{equation}
v_{\rm ej}(t)=\frac{dr_{\rm ej}(t)}{dt}=b \frac{r_{\rm ej}(t)}{t} 
\end{equation}
is the expansion velocity of the early-SN material, where we have used $r_{\rm ej}(t)=r_{\rm em}(t)$, given by Eq.~(\ref{r}).

We have already mentioned that the power-law component in the spectrum of Episode 1 might be due to the accretion onto the NS companion. As this power-law component is present since the beginning of Episode 1, we have fixed the value of $t_{0, \rm accr}$ to be equal to the starting time of Episode 1. This puts a constraint in the separation distance $a$ of the binary, which under these conditions is given by

\begin{equation}\label{a}
a=r_0+R_{\rm cap}(0),
\end{equation}
where $r_0=r_{ej}(0)$ and $R_{\rm cap}(0)$ are the radius of the early-SN ejecta and the capture radius of the NS companion at the beginning of Episode 1. In this case, $r_0 \approx 1.75 \times 10^9$ cm, see Fig. \ref{Rem}. The separation $a$ is a function of the initial mass of the NS and of the SN core progenitor mass, as well as of the orbital velocity, through $R_{\rm cap}$. It is clear that the constraint given by Eq. (\ref{a}) is a lower limit, since the accretion process onto the NS could have been triggered before by layers at lower densities (e.g. He). In such a case, the binary separation $a$ could be higher.

In addition to this constraint, we must take into account that the NS must reach its critical mass $M_{\rm crit}$ at the beginning of Episode 2, since by that time the NS must collapse to a BH and emit the canonical GRB. This implies that  
\begin{equation}\label{taccr}
\Delta t_{\rm accr} \approx \frac{611}{(1+z)} \approx 349 \,\, {\rm s}.
\end{equation} 

We show in Table \ref{Tablaprog} the parameters of the binary system leading to IGC of the NS in a time interval equal to the duration of Episode 1. We adopt an initial mass for the NS, $M_{\rm NS}(0)=1.4 M_{\odot}$ and, correspondingly, a NS radius of $R_{\rm NS}(0)=12.3$ km from the mass-radius relation of \citet{Belvedere}. From the constraint given by Eq. (\ref{a}) we fix the binary separation $a$. We then proceed with the numerical integration of the accretion rate equations by requiring that $M_{\rm NS}(t)=M_{\rm crit}$ at $t=\Delta t_{\rm accr}$, given by Eq. (\ref{taccr}), from which we obtain the efficiency $\eta_{\rm accr}$. 

\begin{table*}
\centering
\begin{tabular}{ccccccccc}
\hline
\hline
$M_{\rm prog}/M_\odot$ & $M_{\rm ej}(0)/M_\odot$ & $\rho_{\rm ej}(0)$ (g cm$^{-3}$) &$\eta_{\rm accr}$ &$\Delta M_{\rm accr}/M_{\rm ej}(0)$&$P$ (min)& $v_{\rm orb,NS}(0)$ (km s$^{-1}$)& $\Delta t_{\rm accr}/P$ & $a/R_{\odot}$\\ 
\hline
4 & 2.7 & $2.39\times 10^5$ & $0.92$ &0.47& 0.52 & $5.24 \times 10^3$& 11.14 &0.037 \\
5 & 3.7 & $3.27\times 10^5$ &  $0.88$ &0.34& 0.45 & $5.84 \times 10^3$ & 12.96  &0.036 \\
6 & 4.7 & $4.16\times 10^5$ & $0.88$& 0.27&0.39 & $ 6.39 \times 10^3$& 14.71&0.035\\
7 & 5.7 & $5.04\times 10^5$ &  $0.89$& 0.22&0.35 &$6.91 \times 10^3$ & 16.39 &0.034 \\
8 & 6.7 & $5.93\times 10^5$ &  $0.91$ & 0.19&0.32 & $7.40 \times 10^3$& 18.00 &0.033\\
9 & 7.7 & $6.81\times 10^5$ &  $0.94$ & 0.16&0.30 & $7.87 \times 10^3$& 19.55 &0.032 \\
10 & 8.7 & $7.69\times 10^5$ &  $0.96$ &0.15& 0.27& $8.32 \times 10^3$& 21.04&0.031  \\
\hline
\end{tabular}
\caption{The massive star - neutron star binary progenitor of GRB 110709B. $M_{\rm prog}$ is the mass of the massive star (in solar masses), $M_{\rm ej}(0)$ is the mass of the ejected material in the early-SN phase (in solar masses), $\rho_{\rm ej}(0)$ is the density of the ejecta at the beginning of the expansion, $\eta _{\rm accr}$ is the efficiency of the accretion process onto the NS, $\Delta M_{\rm accr}=M_{\rm crit} - M_{\rm NS}(0)$ is the total mass accreted by the NS before the collapse, $P=2 \pi a/v_{\rm orb,NS}$ is the period of the binary, $v_{\rm orb,NS}(0)$ is the initial orbital velocity of the NS and $\Delta t_{\rm accr}/P$ is the arc-length travelled by the NS during the accretion process in units of the length of the whole orbit and $a/R_{\odot}$ is the binary separation (in units of solar radii). We suppose that the accretion process starts 5 s before the first trigger, i.e. $t_{0,\rm accr}$ coincides with the time corresponding to the first datapoint in Fig. \ref{broken powerlaw}.}  
\label{Tablaprog}
\end{table*} 

It is interesting to analyze how the NS can accrete such a large mass, in some cases of the order of 47\% of the early-SN material (see column 5 of Table \ref{Tablaprog}), since one could think that solid angles of $\sim 50\%$ between the early-SN material and the accreting NS are hard to obtain. Indeed, during the accretion process the NS is moving with high orbital velocities of the order of $10^8$ cm s$^{-1}$ relative to the core progenitor (see column 7 of Table \ref{Tablaprog}), and consequently travels effective arc-lengths several times larger than the circumference of the orbit (see column 8 of Table \ref{Tablaprog}).

Assuming that the gain in gravitational energy of the accreted material into the NS can be released from the system leads to an upper limit of the luminosity 

\begin{equation}
|\dot{E}_b(t)|=\frac{G \dot{M}_{\rm accr}(t) M_{\rm NS}(t)}{R_{\rm NS}(t)},
\end{equation}
where we take into account the dependence of the NS radius with time, due to the increment of the NS mass by the accretion process. The self-consistent radius is computed at each time from the mass-radius relation of \citet{Belvedere}.

The actual luminosity depends on the efficiency $\eta_{\rm rad}$ in converting gravitational energy into electromagnetic energy by some still unknown process. Since in our model we assume that the BB component of Episode 1 is due to the early-SN expansion, we estimate the efficiency $\eta_{\rm rad}$ from the assumption that $|\dot{E}_b|$ is  responsible for the power-law luminosity $L_{\rm PL}$, namely
\begin{equation}\label{etarad}
\eta_{\rm rad}(t)=\frac{L_{\rm PL}}{|\dot{E}_b(t)|}.
\end{equation}

\begin{figure}
\centering
\includegraphics[width=0.9\hsize, angle=0]{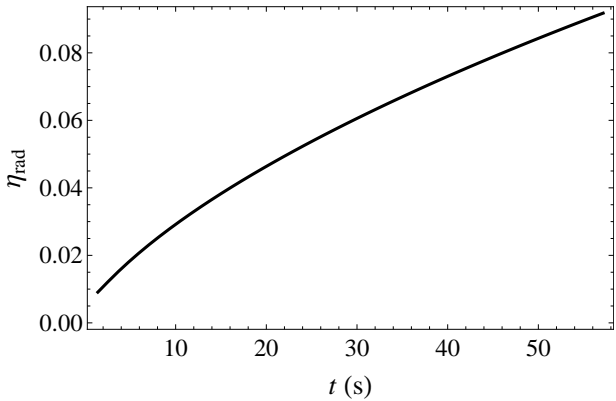}
\caption{Theoretical estimation of the efficiency $\eta_{\rm rad}$ given by Eq. (\ref{etarad}) of the process to convert gravitational energy in radiation as a function of time. For this plot, we have assumed a constant and isotropic power-law luminosity of Episode 1, $L_{\rm PL} \approx 1.8 \times 10^{50}$ erg s$^{-1} \approx 10^{-4}$ $M_{\odot}$ s$^{-1}$. We computed the values of the efficiency for the binary systems shown in Table \ref{Tablaprog}. For all the cases, we obtain the same evolution of the efficiency with time, i.e. the curves overlap. The values of $\eta_{\rm rad}$ are always $< 10 \%$.}
\label{efficiency}
\end{figure}

In Fig. \ref{efficiency} we show the evolution of the efficiency $\eta_{\rm rad}$ in the first seconds of emission for the binary systems shown in Table \ref{Tablaprog}. We assume a constant and isotropic power-law luminosity of Episode 1, $L_{\rm PL} \approx 1.8 \times 10^{50}$ erg s$^{-1} \approx 10^{-4}$ $M_{\odot}$ s$^{-1}$, as found from the spectral analysis. For all the cases, we obtain the same evolution of the efficiency with time, i.e. the curves overlap. This is due to the fact that we constrained all the systems to have the same initial NS mass and $\Delta t_{\rm accr}$ .

\section{Radio observations}\label{radio}

\begin{figure}
\centering
\includegraphics[width=0.9\hsize, angle=0]{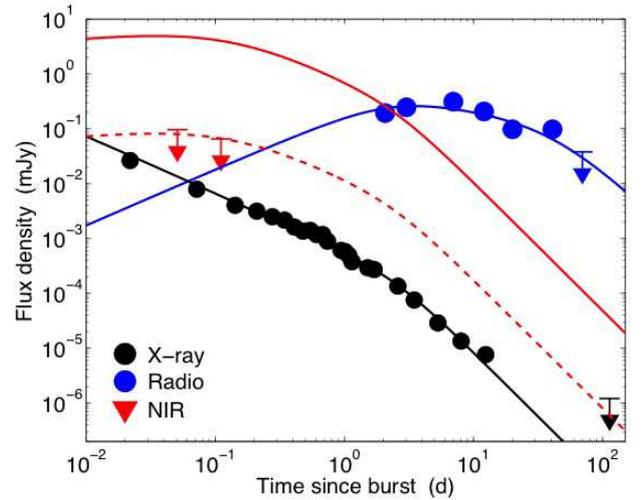}
\caption{X-Ray (black), radio (blue) and NIR (red, upper limits) light curves of GRB 110709B. Taken from \citet{Zaudereretal2012} with kind permission.}
\label{SNradio2}
\end{figure}

\begin{figure}
\centering
\includegraphics[width=0.9\hsize, angle=0]{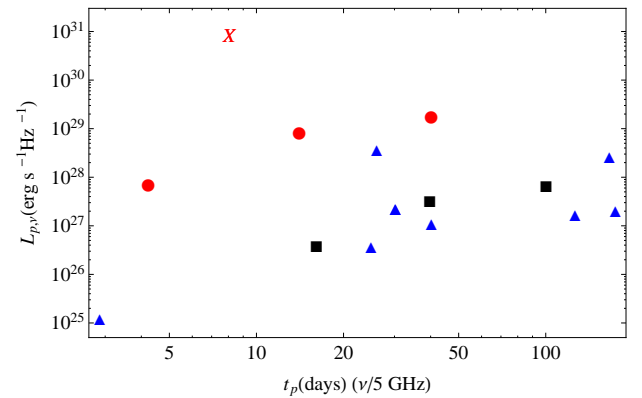}
\caption{Plot of the peak spectral radio Luminosity per unit frequency versus the time at which the peak occurs, for different SN associated to GRBs. The circles represent the SN emission associated to SN 2006aj (GRB 060218), SN 1998bw (GRB 980425) and SN 2003lw (GRB 031203). The triangles represent the SN Ib/c for which there are radio observations, namely SN 2002ap, SN 1990B, SN 2008D, SN 1994I, SN 2009bb and SN 2003L. The squares represent the SN IIb: SN 2008ax, SN 2001ig, SN 1993J, SN 2001gd and SN 2003bg. The red cross is the luminosity related to GRB 110709B afterglow. It is higher than the emissions of the other SN, considered ``standard''.}
\label{SNradio}
\end{figure}

\citet{GCN12190} report observations with the EVLA radio telescopes on several occasions between 11 and 16 July, at a frequency of 5.8 GHz. They found a radio source which brightened by about a factor of 1.6, between 2.1 and 7 days after the burst. The coincidence with the XRT position and the rising flux indicate that this is the radio afterglow of GRB 110709B. The position of the source is RA = 10:58:37.114, DEC = -23:27:16.760. We show in Fig. \ref{SNradio2} the 5.8 GHz light curve presented in \citet{Zaudereretal2012}, where there is evidence of a radio bump. Following the work of \citet{Chevalier}, we have reproduced the plot of the peak spectral radio Luminosity per unit frequency as a function of time (days) at which the peak is produced for different SN associated with GRBs, including GRB 110709B (see Fig. \ref{SNradio}). We find that the radio emission of this source is higher than the ones associated with typical SN.

\section{Conclusions}\label{conclusions} 

GRB 110709B is a very peculiar source, since it is the first for which Swift BAT has triggered twice. Its \textit{Swift} BAT light curve presents two well defined episodes, Episode 1 and Episode 2. Episode 1 lasts 100 s and Episode 2 lasts 135 s. Particularly interesting is the fact that the X-ray observations started well before the second trigger. The light curve and spectrum of this source share similar characteristics with GRB 090618 \citep{LucayJorge}, GRB 101023 \citep{Penacchioni2012} and GRB 970828 \citep{970828}. It has been recently shown that these GRBs which show such distinct emissions, Episodes 1 and 2, form a new family of GRBs described by the IGC paradigm \citep{Jorge,LucayJorge}. Within this scenario, the GRB originates in a binary system formed by a massive star on the verge of a SN and a NS close to its critical mass for the gravitational collapse to a BH. The compact core SN progenitor ejects material in the very early phases of the SN explosion, that is then accreted by the NS; this process is identified with Episode 1. The accretion process onto the NS brings it to the critical mass, leading to its gravitational collapse to a BH and emitting the GRB, identified with Episode 2. Later on, we see a standard emission in X-rays, which we have called Episode 3. Several days after the burst, when it is present, we see an optical emission, associated to the SN (Episode 4). Following the recent works on GRB 090618 \citep{LucayJorge} and GRB 970828 \citep{970828}, we here apply the IGC paradigm to GRB 110709B. 

The redshift of GRB 110709B is unknown, so in Section \ref{redshift estimation} we used four phenomenological methods to constrain it; i.e., Grupe \citep{Grupe}, Amati \citep{Amati(2006)}, Yonetoku \citep{Yonetoku2004} and the scaling of the X-Ray afterglow \citep{Izzo, Penacchioni2012}. The first method gives an upper limit of $z < 1.35$. The second and third methods give a lower limit of $z > 0.6$ and $z > 0.7$, respectively. The last method gives a precise value of $z=0.75$, which lies within the range determined by three the above mentioned methods. We then fixed this last value as the redshift of GRB 110709B.

The spectral analysis of Episode 1 is given in Section \ref{subsec:ep1} and \ref{radius}. We find a value of the isotropic energy for Episode 1 of $E_{iso}^{(1)}= 1.42 \times 10^{53}$ erg (see Table \ref{TABLATODO}). We fit the spectrum with a BB+PL model. The temperature of the BB component evolves with time following a broken power-law (see Fig. \ref{broken powerlaw}). The corresponding radius of the BB emitter evolves in time following a power-law given in Eq. (\ref{r}) and shown in Fig. \ref{Rem}. We associate this radius and the BB component to the evolution of the SN ejecta, while the power-law is associated to the accretion of the ejected material onto the NS companion.

Episode 2 is analyzed in Sections \ref{subsec:ep2} and \ref{episode2}. We find an isotropic energy of  $E_{iso}^{(2)}=2.43 \times 10^{52}$ erg (see Table \ref{P-GRBEiso}). We interpret this episode as a canonical GRB and simulated its light curve and spectrum within the Fireshell model. We find at transparency a Lorentz factor $\Gamma \sim 1.73 \times 10^2$, laboratory radius of $6.04 \times 10^{13}$ cm, P-GRB observed temperature $kT_{P-GRB}= 12.36$ keV, baryon load $B=5.7 \times 10^{-3}$, P-GRB energy of $E_{P-GRB}=3.44 \times 10^{50}$ erg, and a CBM mean density $76$ part cm$^{-3}$. This value is consistent with a ``dark GRB'', as cited in \citet{Zaudereretal2012}. The lack of detection of a SN emission for this particular GRB could be due to obscuration by the circumstellar dust in the host galaxy.  

The nature of the progenitor is discussed in Section \ref{Nature of the Progenitor}.  We indicate that it is a binary system formed by a massive evolved star on the verge of a SN explosion and a NS. We associate the thermal component of Episode 1 mainly with the early-SN evolution and the power-law component to the accretion process onto the NS. There is the possibility that also the accretion process has a thermal contribution. The energy due just to the thermal component is of the order of $10^{50}$ erg, which is reasonable for the expansion of the early-SN ejecta. We perform all the necessary calculations to obtain the parameters of the binary system. For all our calculations we assume a fixed NS mass of $1.4 M_{\odot}$. We compute the rate at which the early-SN material enters the capture region, for given values of the SN core progenitor mass. From this material, only a fraction will be accreted by the NS, so we introduced an efficiency factor $\eta_{\rm accr}$.  As the power-law component is present since the beginning of Episode 1, we suppose that this episode starts at the same time $t_{0, \rm accr}$ as the accretion process, namely, when the outermost shell of expanding ejecta reaches the capture radius $R_{\rm cap}$ of the NS (measured from the center of the NS). This puts a constraint on the separation distance $a$ of the binary. In addition, the NS must reach its critical mass and collapse to a BH at the beginning of Episode 2. This puts a constraint on the duration of the accretion process $\Delta t_{\rm accr}$. By integrating the accretion rate equations with these boundary conditions we obtain the efficiency $\eta_{\rm accr}$. We summarize the results in Table \ref{Tablaprog}, for different values of the core-progenitor mass and the density of the early-SN ejecta. Assuming that the power-law radiation comes from the conversion of the binding energy of the accreted material onto the NS, we estimate the efficiency $\eta_{\rm rad}$ of this conversion process, which we show in Fig. \ref{efficiency} for an isotropic power-law luminosity $L_{\rm PL} \approx 1.8 \times 10^{50}$ erg s$^{-1} \approx 10^{-4}$ $M_{\odot}$ s$^{-1}$ observed in Episode 1. For the parameters of the binary system shown in Table \ref{Tablaprog}, we obtain values of $\eta_{\rm rad} < 10 \%$. The efficiency of the radiation mechanism can be even lower if some beaming or boosting is present. However, we did not address any such possible mechanism in this work.

In section \ref{radio} we present the radio observations of GRB 110709B with the EVLA radio telescopes and the X-ray, radio and NIR light curves taken from \citet{Zaudereretal2012}. We  notice the presence of a bump in the radio afterglow, at $\approx 10$ days after the burst. As GRB 110709B has been classified as an optically dark burst, we plotted the peak spectral radio luminosity per unit frequency as a function of time and compared it with the luminosities of typical SNe, to see if it was possible to find any coincidences that may indicate the presence of the SN in the radio band. However, the luminosity we find is much higher than the ones of the standard SNe. 

We interpret, within the IGC paradigm, that GRB 110709B is a new member of the IGC family, in addition to GRB 090618, GRB 101023 and GRB 970828. 

A remarkable support of the above IGC paradigm comes from the observations of the X-ray afterglow emission of these systems. The X-Ray light curve is composed of an early steep decay, a plateau and a late decay. The analysis of the late decay of the afterglow luminosity has been identified with the cooling of the newly born NS, left by the SN explosion \citep{Negreiros}. 

\begin{acknowledgements}

We are very grateful to the anonymous referee for his/her comments and suggestions that helped to improve the presentation of our results. We thank the \textit{Swift} team for the support. This work made use of data supplied by the UK \textit{Swift} Data Centre at the University of Leicester. A. V. P. is supported by the Erasmus Mundus Joint Doctorate Program by Grant Number  2010-1816 from  the EACEA   of the European Commission. G. B. P. is supported by the Erasmus Mundus Joint Doctorate Program by Grant Number  2011-1640 from  the EACEA   of the European Commission.

\end{acknowledgements}

\end{document}